\documentclass[12pt]{spieman}  % 12pt font required by SPIE;
\usepackage{amsmath,amsfonts,amssymb, gensymb}
\usepackage{graphicx, caption, subcaption}
\usepackage{tabu, tabularx}
\usepackage{setspace}
\usepackage{tocloft}
\usepackage{lineno}
\title{Fast and Scalable Production of Stacked Prism X-ray Lenses for Astrophysics Using Two-Photon Polymerization}

\author[a,b*]{Filip af Malmborg}
\author[c]{Chloé Delmotte}
\author[d]{Kian Shaker}
\author[a,b]{Mark Pearce}
\affil[a]{KTH Royal Institute of Technology, Department of Physics, SE-106 91 Stockholm, Sweden}
\affil[b]{The Oskar Klein Centre for Cosmoparticle Physics, AlbaNova University Center, SE-106 91 Stockholm, Sweden}
\affil[c]{KTH Royal Institute of Technology, Department of Intelligent Systems, SE-100 44 Stockholm, Sweden}
\affil[d]{KTH Royal Institute of Technology, Department of Applied Physics, SE-106 91 Stockholm, Sweden}

\cftpagenumbersoff{figure}
\cftpagenumbersoff{table} 
\begin{document} 
\maketitle

\begin{abstract}
Stacked prism lenses (SPLs) are a type of refractive X-ray optics currently under development with the potential to greatly improve on current X-ray telescope optics in terms of focal length, angular resolution, efficiency and scalability. For this work, SPLs are manufactured using two-photon polymerization (2PP), with production being significantly faster and with higher geometric fidelity than previous methods. Preliminary laboratory tests show improved efficiency compared to previous manufacturing methods and promising optical capabilities. Two-photon polymerization is shown to be a reliable method for producing SPLs, and when challenges around printing time and assembly are addressed, the path towards an SPL X-ray telescope lies open.
\end{abstract}

% Include a list of up to six keywords after the abstract

% Include email contact information for corresponding author
{\noindent \footnotesize\textbf{*}Filip af Malmborg,  \linkable{filipam@kth.se} }

\begin{spacing}{2}   % use double spacing for rest of manuscript

\section{Introduction}

X-ray optics in the form of grazing incidence mirror assemblies have dominated X-ray astronomy since the launch of the first orbiting X-ray telescope, {\it Einstein/HEAO2}, in 1978\cite{giacconi_einstein_1979}. Wolter type-1 mirrors\cite{wolter1952spiegelsysteme}, as used on {\it Einstein/HEAO2}, are also used on contemporary missions such as \textit{Chandra}, \textit{XMM-Newton} and \textit{XRISM}. While proven technology, these mirrors have long focal lengths ($\sim 10$~m), high weight ($\sim 500$~kg) and either poor ($>10$~arcseconds) angular resolution (\textit{XMM, XRISM}) or relatively small ($\sim 500$~cm$^2$) effective area (\textit{Chandra}).

Grazing-incidence optics are also proposed for the next generation of X-ray telescopes, which aim to combine arcsecond angular resolution with a large effective area through low mass optics\cite{christensen_x-ray_2022}. With a planned launch date of 2037, the ESA {\it NewAthena} mission (0.1--12~keV) adopts a development of Wolter-I optics, Wolter-Schwarzschild optics, yielding a flatter focal surface and reduced spherical aberration and coma. X-rays are reflected within channels formed between stacked silicon wafers, known as silicon pore optics, SPO\cite{collon_development_2022}. The 2.5~m diameter mirror comprises 90\,000 wafer elements robotically stacked to form 600 individual SPO imagers arranged in concentric annuli. The focal length is 12~m and the angular resolution 5 arcsec. A similar concept called micro-pore optics (MPO) is used on the in-orbit Einstein Probe (a joint CSA-ESA mission) for its Wide-field X-ray Telescope (WXT)\cite{yuan_einstein_2018}. The US-led {\it Lynx} mission (0.2--10~keV) aims to achieve a 0.5 arcsec angular resolution (similar to {\it Chandra}), but with a 20-fold improvement in effective area and improved off-axis response. In order to meet launcher mass constraints, this requires the development of thin reflectors, which are polished to meet figure/surface roughness requirements. Another approach being considered is an assembly of thin reflectors paired with piezoelectric actuators, allowing the figure to be actively adjusted. Such an assembly comprises $\sim$40\,000 reflectors\cite{zhang_high-resolution_2019}.
These future missions face formidable assembly challenges, which will necessitate a modular approach with automated fabrication procedures - a theme which also characterizes the work presented in this paper.

In this paper, we describe an alternative approach for an X-ray telescope based on {\it refractive} optics. 
For most materials, the X-ray refractive index, $n$, is close to unity ($n \lesssim 1$), which combined with significant attenuation constitutes a major design challenge. The first demonstration of refractive X-ray optics at a synchrotron light source was made in 1996\cite{snigirev_compound_1996}.  Since $n<1$, focusing requires a concave lens. To allow a feasible radius of curvature of the lens surface and adequate aperture, many lenses were stacked together. Such a compound refractive lens (CRL) was fabricated by drilling 30 co-aligned holes (600~µm diameter, 25~µm separation) into a Al-Cu alloy block.     
Now-a-days, CRL assemblies are routinely used in X-ray microscopy set-ups at synchrotron light sources\cite{kohn_diffraction_2003}. A low-atomic number element such as beryllium or carbon is often used for the lens material since the mass attenuation coefficient increases rapidly with atomic number. To mitigate spherical aberration and to maximize the aperture, parabolic lens surfaces are used\cite{lengeler_parabolic_2002}. An important issue is that the refractive index of the lens material, and consequently the focal length of the lens depends on the X-ray energy. For monochromatic synchrotron light source applications, this is not a problem. For the broadband spectra in astronomy, this is an important effect to address. 

\begin{figure}
    \centering
    \includegraphics[width=0.9\linewidth]{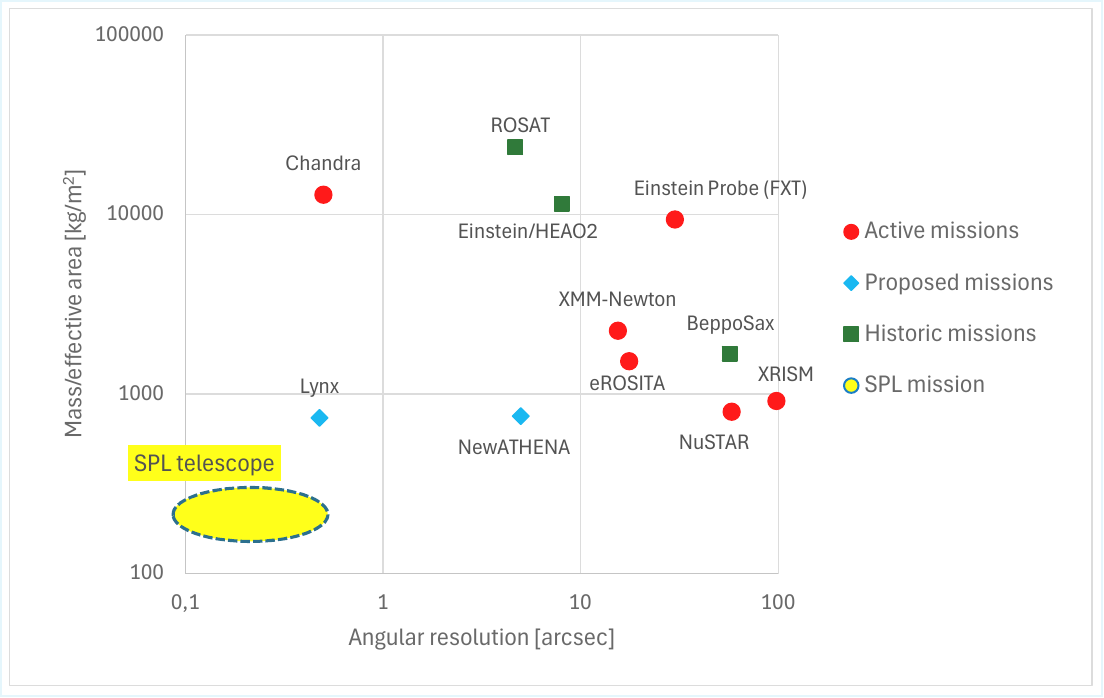}
    \caption{Comparison of the mass/effective area ratio and angular resolution for a future SPL telescope with existing and proposed X-ray telescope designs. Adapted from Ref.~\citenum{christensen_x-ray_2022}, with data for proposed SPL telescope from Ref.~\citenum{mi_stacked_2019}.}
    \label{fig:telescopecomparison}
\end{figure}

Refractive optics have had no significant role in X-ray astronomy. Diffractive-refractive optics, in the form of Fresnel zone plates, have been studied, but the extremely long focal length (up to 10$^3$~km!) requires that the optics and detectors are distributed between formation-flying spacecraft - an approach which is currently not feasible\cite{gorenstein2004role}. 
In this paper, we evaluate the performance of a new type of refractive X-ray optics, the Stacked Prism Lens (SPL) \cite{mi_stacked_2019}.
The lens material is arranged in a replicating prism geometry to approximate a parabolic lens\cite{jark_focusing_2004}. The SPL is characterized by a short focal length ($\lesssim$1~m), high efficiency and high angular resolution (potentially sub-arcsecond). The low weight and inherent scalability of an SPL assembly make it an attractive solution for an X-ray telescope mission, albeit with potentially challenging tradeoffs owing to the inherent chromaticity of refractive optics. A comparison of possible capabilities of an SPL telescope with current and upcoming X-ray telescopes can be found in Figure~\ref{fig:telescopecomparison}.

Previously, prism-array lenses such as SPLs have been manufactured by rolling a ridged polyamide film\cite{nillius_large-aperture_2011} or focused UV lithography on SU-8 photoresist\cite{mi_stacked_2019}, both of which have significant drawbacks in terms of production speed and geometric precision. The UV lithography lens was tested at the Diamond X-ray synchrotron light facility and showed promising optical capabilities, but with suboptimal angular resolution due to manufacturing defects. 

In this paper, we show how SPLs can be manufactured from epoxy-like polymers by two-photon polymerization (2PP)\cite{maruo_three-dimensional_1997} using a commercially available process. We show that the 2PP-manufactured lenses improve greatly on previous methods in terms of desired geometry and manufacturing speed, meet theoretical predictions in terms of efficiency and produce a focal spot smaller than the resolution and spot sizes of the laboratory equipment available for this work.

This paper is arranged as follows. In Section~\ref{sec:SPL_design}, the concept and design parameters of an SPL is introduced, followed by Section~\ref{sec:2pp}, where the two-photon polymerization manufacturing method is explained. Then, Section~\ref{sec:testing_SPLs} provides results from SPL tests, showing that the production method produces working lenses with promising performance. Lastly, Section~\ref{sec:conclusions} addresses some outstanding challenges in the development of the SPL concept for an X-ray telescope, and provides an outlook for future work.

\section{Stacked Prism Lens Design}\label{sec:SPL_design}

\begin{figure}
    \centering
    \includegraphics[width=0.99\linewidth]{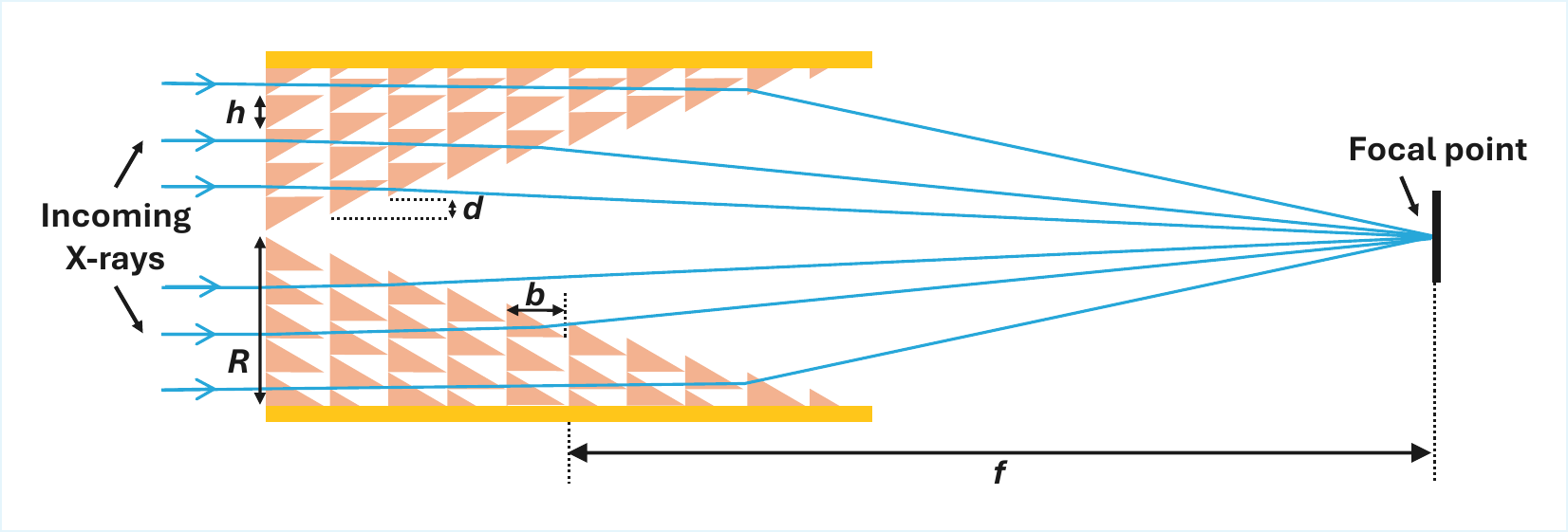}
    \caption{Working principle of an SPL lens. Distances and lengths are not to scale, typical values of the parameters are: $b \sim 50$~µm, $h \sim 10$~µm, $d\sim 3$~µm,  $R \sim 200$~µm, $f \sim 0.3$~m. Figure adapted from Ref.~\citenum{mi_stacked_2019}.}
    \label{fig:spl_principle}
\end{figure}

The design of an SPL has previously been described in Ref.~\citenum{mi_stacked_2019} and Ref.~\citenum{nillius_large-aperture_2011} and so is only briefly introduced here. An SPL lens approximates a parabolic lens with material cut away corresponding to a $2\pi$ phase shift in the light wave, resulting in a series of small ($\sim 10 $~µm) prisms. Compared to the stacked parabolic lenses of CRL arrays, SPLs can achieve significantly shorter focal lengths (order of decimeters rather than meters\cite{simons_simulating_2017}) while keeping an efficiency of over 50\%. A schematic drawing of an SPL can be found in Figure~\ref{fig:spl_principle}. The focal length $f$, radial prism width $h$, prism length $b$, the columnar displacement $d$ and  the deviation from unity of the real part of the refractive index $\delta$ are related as:

\begin{equation}\label{eq:spl_parameters}
    f = \frac{dh}{b\delta},
\end{equation}

\noindent where $b = m\lambda/\delta$ (where $m$ is a whole number and $\lambda$ is the wavelength), to ensure phase continuity. Since $\delta \propto E^{-2}$, the lens is chromatic, and the focal length for a given set of prism dimensions will vary as $f \propto E^2$. Thus, the lens will be designed for a particular energy, and other energies will be out of focus. This is a major drawback of refractive optics for astrophysics, as discussed further in section~\ref{sec:conclusions}. The refractive index deviation $\delta$ of the lens material was calculated from the chemical formula and density of the polymer used for manufacturing, with data from Ref.~\citenum{elam_new_2002}, fetched using the Python {\tt XrayDB} package.

\subsection{Optimizing SPL design}\label{sec:optimizing_spl} 

In addition to the parameters of Equation~\ref{eq:spl_parameters}, the radius $R$ of an SPL provides an additional free parameter. The larger the radius, the more layers of prisms are needed in the optical path to focus the light from the outer edges of the aperture. This increases the length of the lens and decreases the efficiency at the edge of the aperture, since more material attenuates the incoming light. The efficiency depends on the transmission function integrated across SPL radius, shown by Ref.~\citenum{nillius_large-aperture_2011} to be:

\begin{equation}\label{eq:transmission_func}
    \epsilon = \frac{2}{R^2} \int_0^{R} e^{-\frac{b}{2d}\mu -\frac{r^2}{2\delta f^2}} r dr,
\end{equation}
\noindent where $\mu$ is the linear attenuation coefficient and the other parameters are as described in Equation~\ref{eq:spl_parameters}.

\begin{figure}
    \centering
    \includegraphics[width=0.9\linewidth]{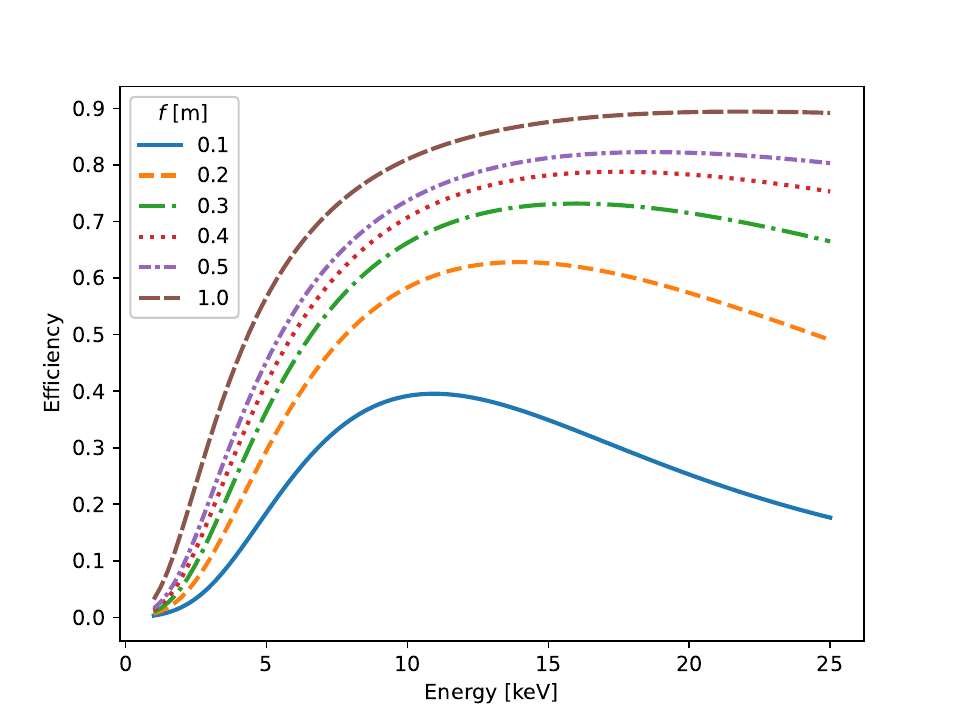}
    \caption{Efficiency of different SPL designs across a range of focal lengths and energies, with a fixed radius of $R = 200$~µm.}
    \label{fig:tradeoff_200µm}
\end{figure}

The SPL design requires a trade-off in terms of focal length, radius and efficiency, which has been studied for this work. In Figure~\ref{fig:tradeoff_200µm}, $R = 200$~µm while the energy and focal length are varied. The figure shows that SPL efficiency drops sharply in the soft X-ray band due to photoelectric absorption. Efficiencies above 50\% can be achieved by choosing a longer focal length, but the efficiency gain is marginal for $f > 0.3$~m.

\begin{figure}
    \centering
    \includegraphics[width=0.9\linewidth]{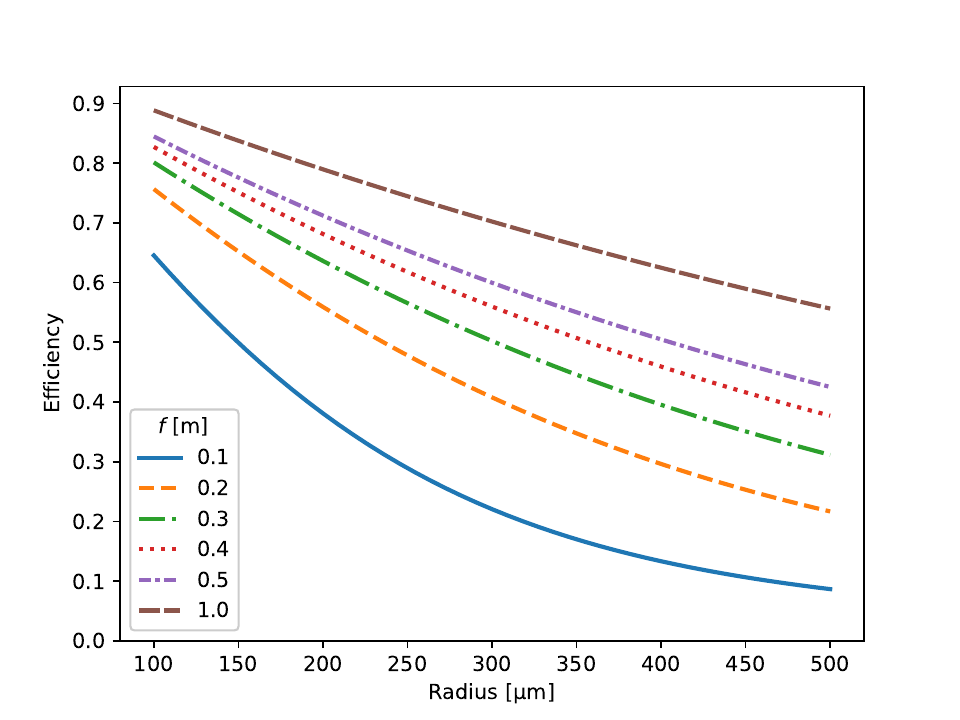}
    \caption{Efficiency of different SPL designs across a range of focal lengths and radii, at a fixed energy of $E = 9.2$~keV.}
    \label{fig:tradeoff_9.2kev}
\end{figure}

The X-ray test setup (further described in Section~\ref{sec:testing_SPLs}) available for this work uses an X-ray source with a prominent emission peak at 9.2~keV. The efficiency for different focal lengths is plotted against the radius at this energy in Figure~\ref{fig:tradeoff_9.2kev}. From the figure, an SPL design with $R = 200$~µm and $f = 0.3$~m was chosen, which gives an expected efficiency of 63.6\%. In the future, this choice of focal length and radius will be guided by constraints and trade-offs in terms of the size and scope of an SPL telescope, manufacturing speed and desired efficiency. For this work, the choice was mainly guided by manufacturing constraints which will be discussed in section~\ref{sec:2pp} as well as test setup constraints, to be discussed in section~\ref{sec:testing_SPLs}. 
The ratio between $h$ (the radial width of the prisms) and $d$ (the layer displacement) is a free parameter in the lens design. A larger ratio means a smoother (linear) approximation of a parabolic lens and therefore better angular resolution, but conversely means that more layers are needed and therefore reduced efficiency. The lens previously manufactured with UV lithography\cite{mi_stacked_2019} used $h=d$ to simplify manufacturing. For this  paper with a more flexible manufacturing method, a ratio of $h = 3d$ was chosen. This choice strikes a reasonable balance between angular resolution and efficiency for the SPL prototyping and development which this work covers. The parameter $d$ and the ratio $h/d$ was discussed in some detail for an earlier iteration of prism-array X-ray lenses \cite{cederstrom_generalized_2005}, and will need to be studied further for SPL telescope design. 

\begin{figure}
    \centering
    \includegraphics[width=0.9\linewidth]{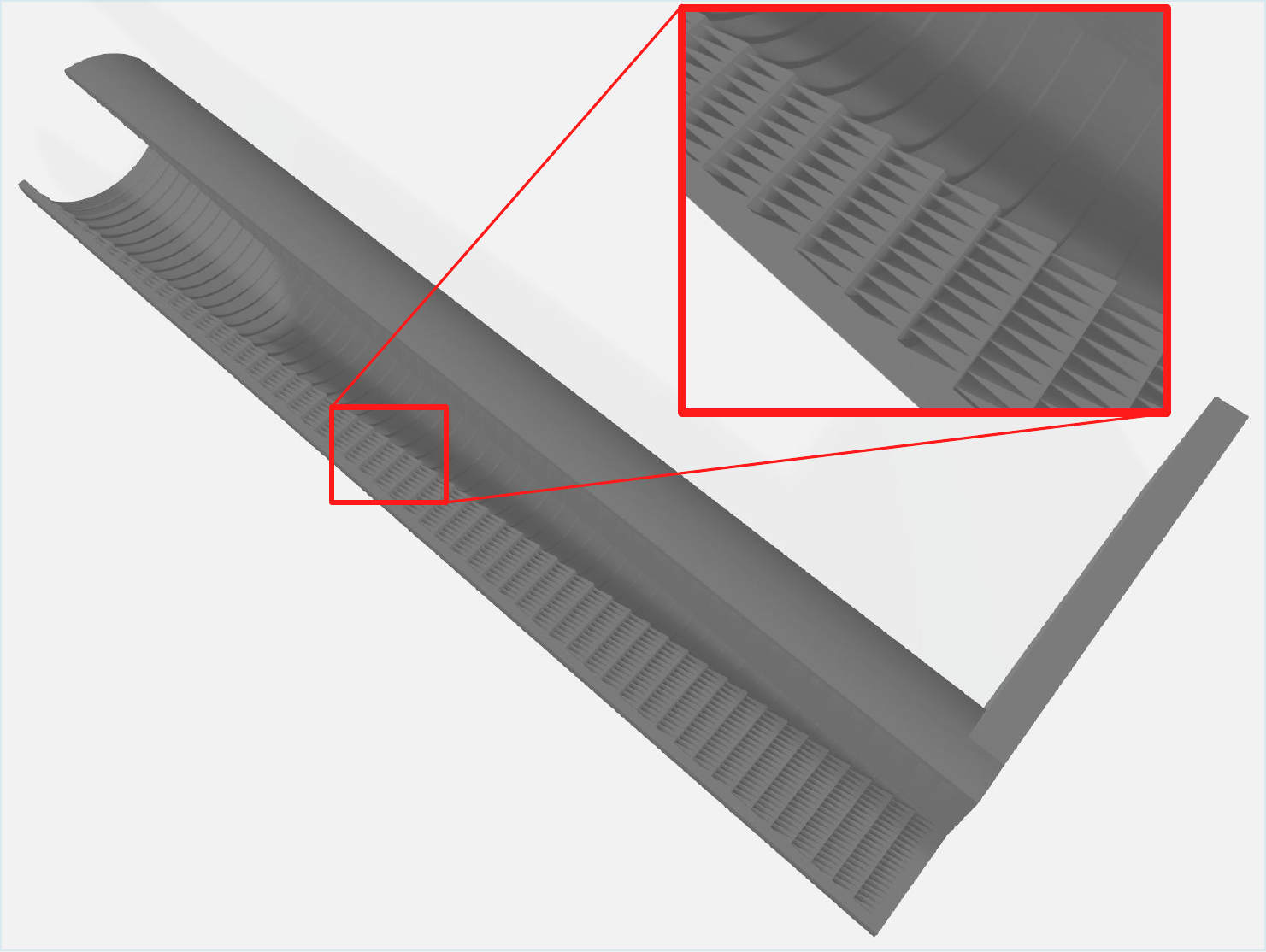}
    \caption{CAD model of final lens design, with a 90 degree slice taken out to show the interior, and a zoomed in area to show the rows of prisms in more detail. The lens has a radius $R = 220$~µm (including the support walls) and a length $L = 2499$~µm.}
    \label{fig:lens_cad}
\end{figure}

The resulting prism parameters for the final design for this work are $b = 43.11$~µm, $h= 11.01$~µm, $d = 3.67$~µm. The diffraction-limited FWHM resolution of this design is 76~nm, corresponding to 0.052~arcseconds. Naturally, this will always be a theoretical highest resolution, and manufacturing deficiencies and material impurities will ensure that the resolution for any real lens is larger than this. For the lens manufactured and tested in Ref.~\citenum{mi_stacked_2019}, there was a 20-fold difference between theoretical and measured FWHM due to manufacturing deficiencies. One of the main drivers of continued SPL development (and the motivation for a new production method as explored in this paper) is to improve on this, with the aim of achieving sub-arcsecond resolution. For this paper, the SPL was produced with a 2~µm layer between each layer of prisms and 20~µm walls surrounding the whole lens structure to improve structural stability. Additionally, a protruding tab was added at the bottom of the lens to simplify handling. All the final SPL parameters can be found in Table~\ref{tab:SPL_parameters}, and a CAD rendering of the final lens design can be found in Figure~\ref{fig:lens_cad}.

\begin{table}[h]
    \centering
    \setlength{\extrarowsep}{4pt}
    \begin{tabu}{|X[0.5,c]|X[c]|X[c]|}
        \hline \hline
        Parameter & Value & Explanation\\
        \hline
        $f$ & 0.3~m & Focal length \\
        $R$ & 200~µm & Lens radius \\
        $E$ & 9.2~keV & Design energy \\
        $b$ & 43.11~µm & Prism length \\
        $h$ & 11.01~µm & Prism width \\
        $d$ & 3.67~µm & Layer-to-layer displacement\\
        $L$ & 2499~µm & Length of lens\\
        $\epsilon$ & 63.6\% & Theoretical efficiency\\
        FWHM & 0.052 arcsec & Diffraction-limited resolution\\
        \hline
    \end{tabu}
    \caption{Parameters for the SPL manufactured for this work.}
    \label{tab:SPL_parameters}
\end{table}

\section{Two-photon polymerization manufacturing}\label{sec:2pp}

\subsection{Principles of two-photon polymerization}

\begin{figure}
    \centering
    \includegraphics[width=0.9\linewidth]{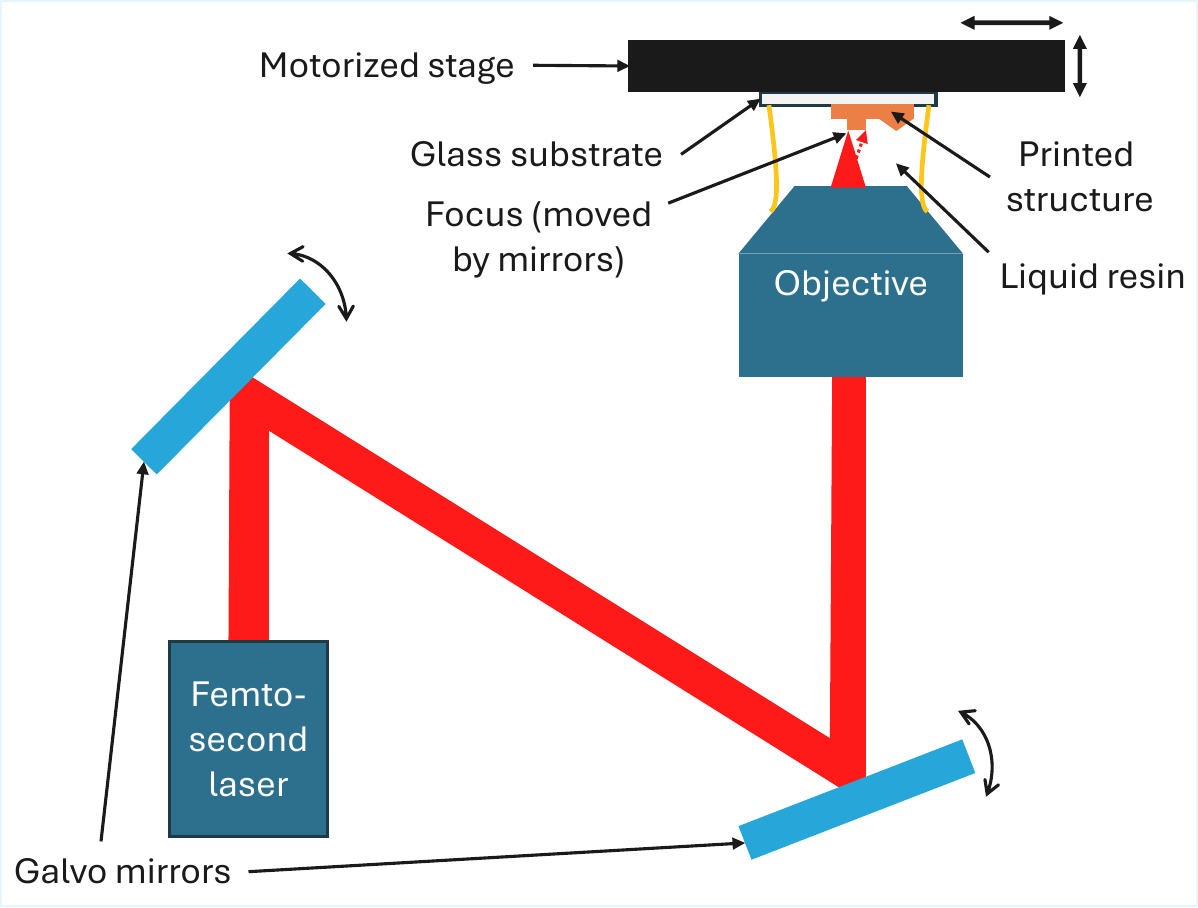}
    \caption{Schematic (not to scale) image of the principle behind 3D printing with two-photon polymerization. Image adapted from Nanoscribe material.}
    \label{fig:2pp}
\end{figure}

3D printing by two-photon polymerization (2PP) was first conceived of in the 1990's \cite{maruo_three-dimensional_1997}. Utilizing a femtosecond-pulsed optical or near-infrared laser to precisely tune the energy dose delivered, a liquid resin is polymerized only when two photons are absorbed\cite{somers_physics_2023}. With a focal spot smaller than 1~µm and the correct pulse frequency, the resin can be hardened in voxels as small as 0.1~µm. By moving this focal spot with mirrors mounted on very precise servos (so-called galvo mirrors) and moving the sample stage with stepper motors, a 2PP 3D printing machine can build up a geometry voxel by voxel. A schematic of this setup can be seen in Figure~\ref{fig:2pp}.

Nanoscribe was the first company to develop a commercial 2PP 3D printing machine in the 2000's. The machine used in this work is a Nanoscribe Photonic Professional GT2. Printing is done by loading a 3D drawing of the desired structure into the job preparation program {\it Describe}, where a printing recipe to set various print parameters can be chosen. This information is then fed into the {\it Nanowrite} program that handles the actual printing process, and a drop of resin is inserted into the machine on a glass substrate. After printing is finished, leftover resin is washed away in propylene glycol methyl ether acetate (PGMEA) solvent, and the print is cleaned by washing it in isopropanol. Then, any leftover isopropanol is dried off using compressed air, and the print can be removed from the substrate.

\subsection{Optimizing the printing process}

For the manufacturing of SPLs, the Nanoscribe resin IP-S (chemical formula: $\mathrm{C}_{14}\mathrm{H}_{18}\mathrm{O}_{7}$, density (after polymerization): 1.2 g/cm$^3$, $\delta = 1.06 \times 10^{-5} - 4.22 \times 10^{-7}$ in the range 5-25~keV) was chosen for its high proximity effect, which means that neighboring voxels are partially polymerized when exposing a voxel. This softens the boundaries between voxels and produces smooth surfaces suitable for micro-optics. The IP-S resin has a refractive index similar to glass, so a glass substrate covered with a thin indium-tin oxide (ITO) layer was used. This provides an interface layer with a change in refractive index, which allows the machine to locate and start printing at the boundary between resin and substrate. A 25x objective (numerical aperture = 0.8) was chosen to allow sub-micron structures to be printed. The objective field-of-view has an effective diameter of $\sim400$~µm. The lens was printed in a standing configuration and the design radius was chosen to be 200~µm, which allowed the whole SPL to fit inside the field of view of the objective. This avoided splitting the lens into several blocks which would produce seams and increase the printing time. Vertically, the 25x objective has a working distance of around 380~µm, so the lens was cut in vertical blocks which align with a whole number of layers, to minimize the effect of the seams. By printing this way, stage movement is kept to a minimum, which optimizes the print accuracy and printing speed.

The print fidelity and printing time was further optimized by performing a parameter sweep of the most important printing parameters: slicing/hatching distance (horizontal and vertical step size, respectively), laser power and scan speed. The latter two determine the size of the voxel being polymerized, and should thus be closely matched to the slicing and hatching distances. This parameter sweep was made printing only a few layers of an SPL to reduce printing time, from several hours to $\sim 20$~minutes. The resulting prints were imaged using a Carl Zeiss Axioscope 5 optical microscope and a Helios 5 UC scanning electron  microscope (SEM). For the SEM study there was an issue with charge-collection on the non-conductive polymer, causing electron beam deflection and image distortion. This was remedied by sputtering a thin layer ($\sim 10$ nm) of carbon nanoparticles on the surface.

\begin{figure}
    \centering
    \includegraphics[width=0.7\columnwidth]{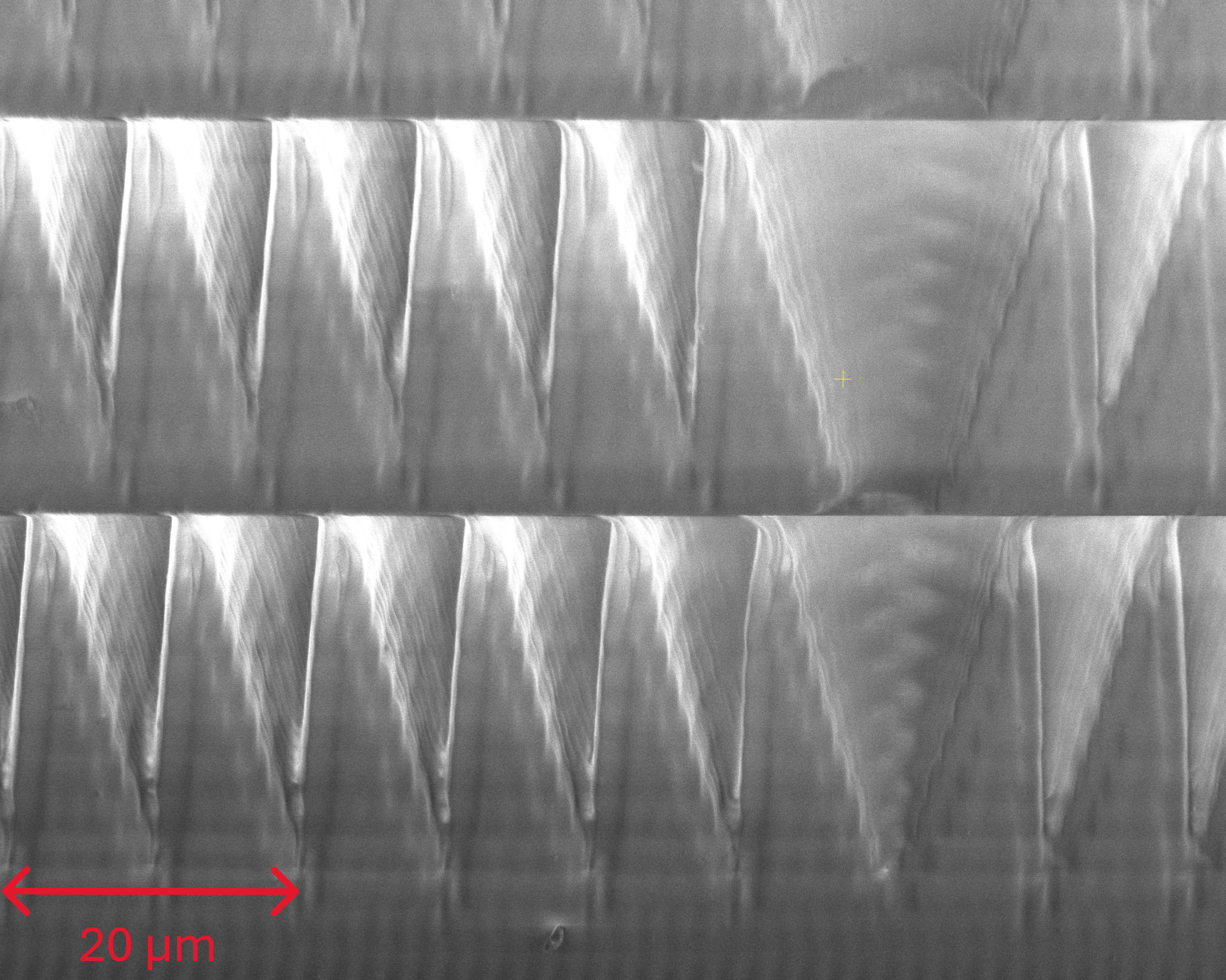}
    \caption{SEM image of rows of SPL prisms from a parameter scan print using a too-coarse slicing distance of 1.0~µm, resulting in unwanted staircase structures.}
    \label{fig:pscan_steps_SEM}
    \end{figure}

\begin{figure}
    \centering
    \includegraphics[width=0.8\linewidth]{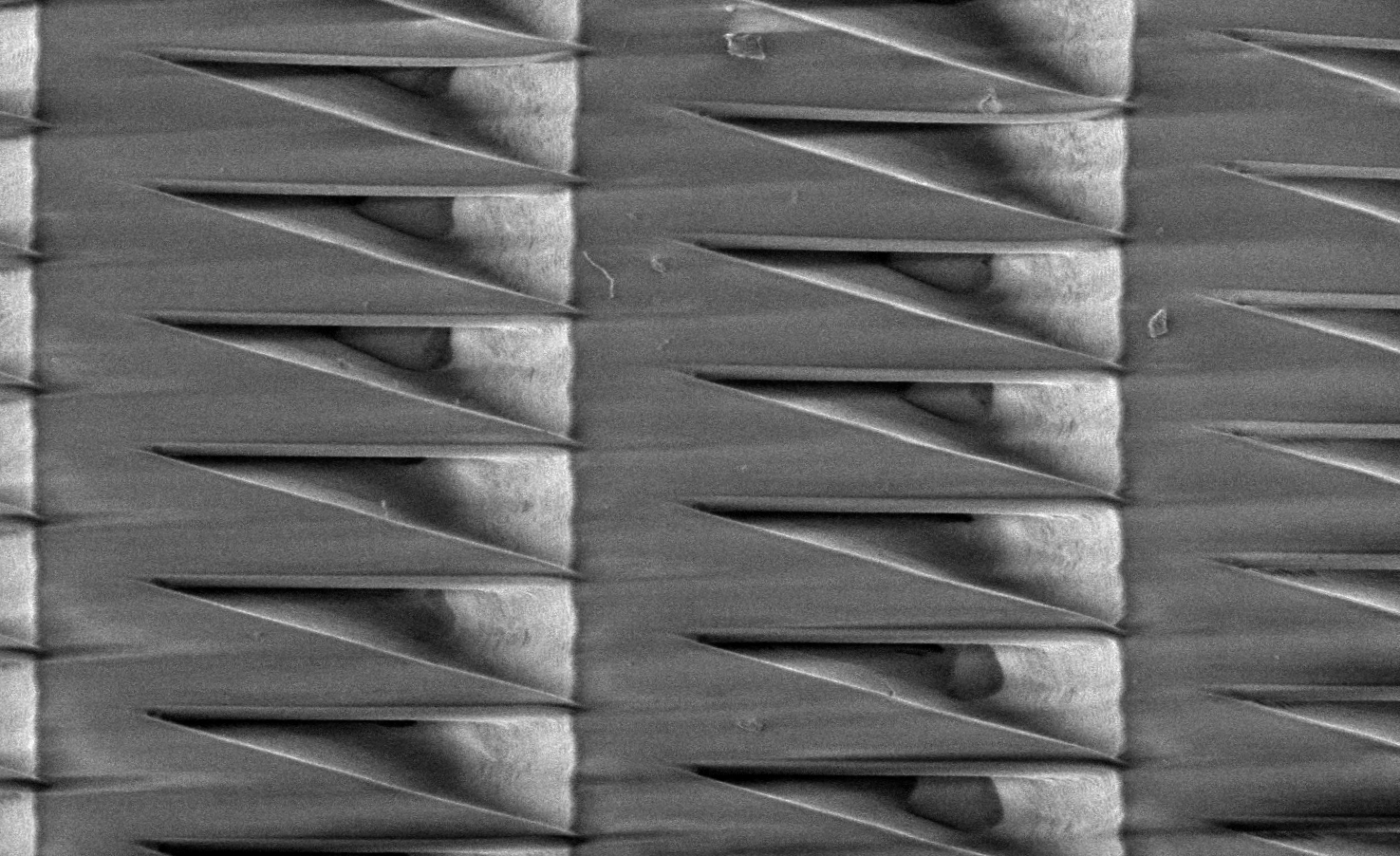}
    \caption{SEM image of print with optimized print parameters, showing a sufficiently smooth surface. All prism tips in this image are thinner than 1~µm which is representative of the general print quality.}
    \label{fig:smooth_triangles}
\end{figure}

\begin{figure}
    \centering
    \includegraphics[width=0.8\linewidth]{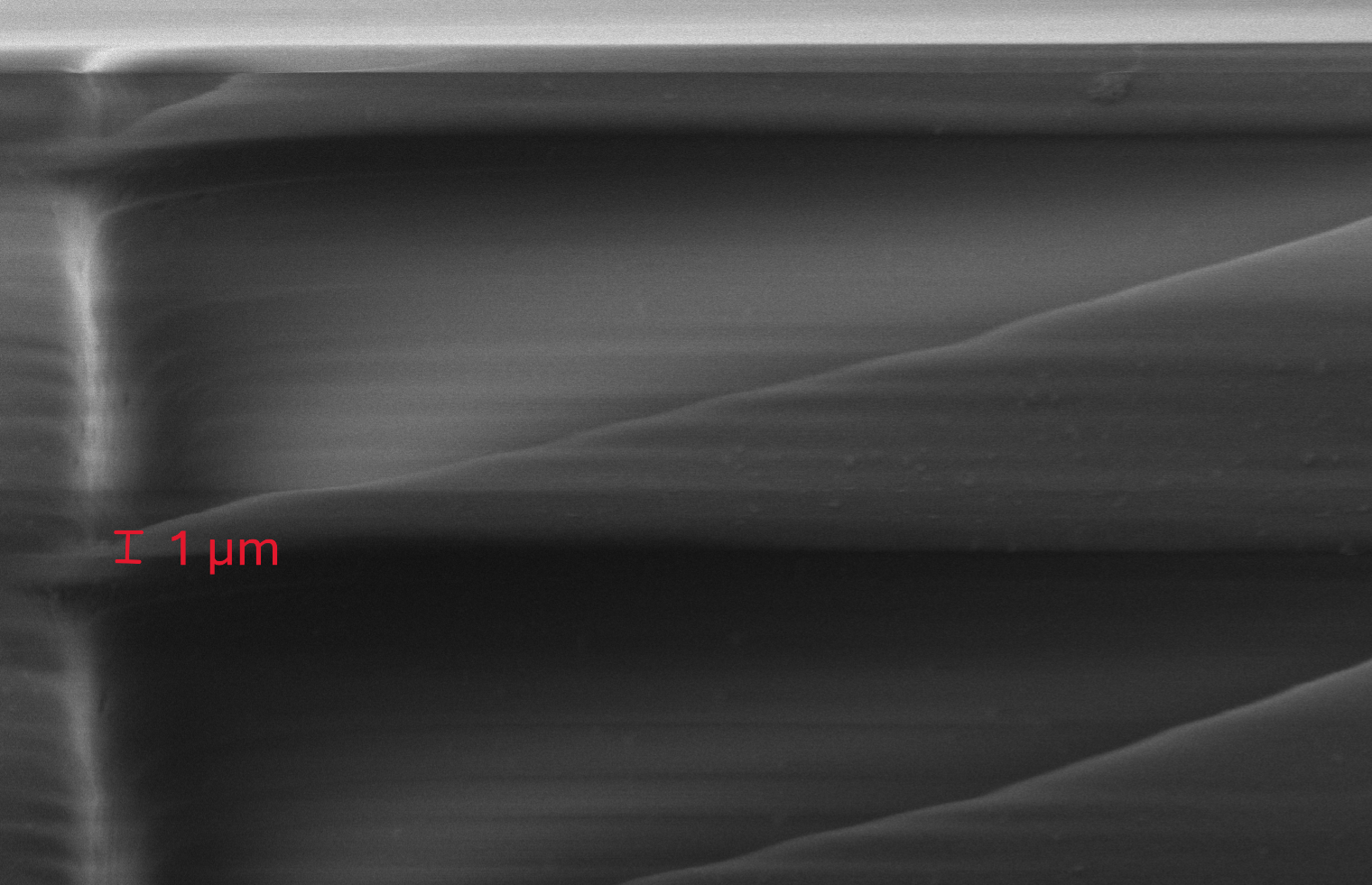}
    \caption{Zoomed-in SEM image of slightly widened tips of prisms. Most prism tips were found to be thinner than the 1~µm shown in the image.}
    \label{fig:thickest_tips}
\end{figure}

Slicing and hatching distance greatly affects the printing speed and should be kept as large as possible; but with too large distances the corresponding voxels were found to produce step-like structures such as those shown in Figure~\ref{fig:pscan_steps_SEM}. The ratio between slicing (horizontal) and hatching (vertical) distances should be the same as the aspect ratio of the critical features, in this case the prisms. Additionally, the polymerization voxels are elongated along the optical axis of the laser. To match this, the SPL was printed in a standing configuration, which aligns the long side of the prisms with the elongated voxels. For the chosen SPL design, the prism aspect ratio is around 3.95:1, and from the parameter scan values of 0.75~µm (slicing) and 0.19~µm (hatching) were found to produce a smooth surface, as can be seen in Figure~\ref{fig:smooth_triangles}. Some defects were found at the tips of the prisms, with some tips having width on the order of 1-2~µm, as seen in Figure~\ref{fig:thickest_tips}. This is an improvement on the method used by Ref. \citenum{mi_stacked_2019}, where the tips of all prisms had a width of $\sim 5$~µm and several prisms were broken or deformed during manufacturing and assembly. Across the surfaces of the whole lens, general irregularities were smaller than 100 nm, with no bumps or pits larger than 1~µm found.

\begin{figure}
    \centering
    \includegraphics[width=0.7\columnwidth]{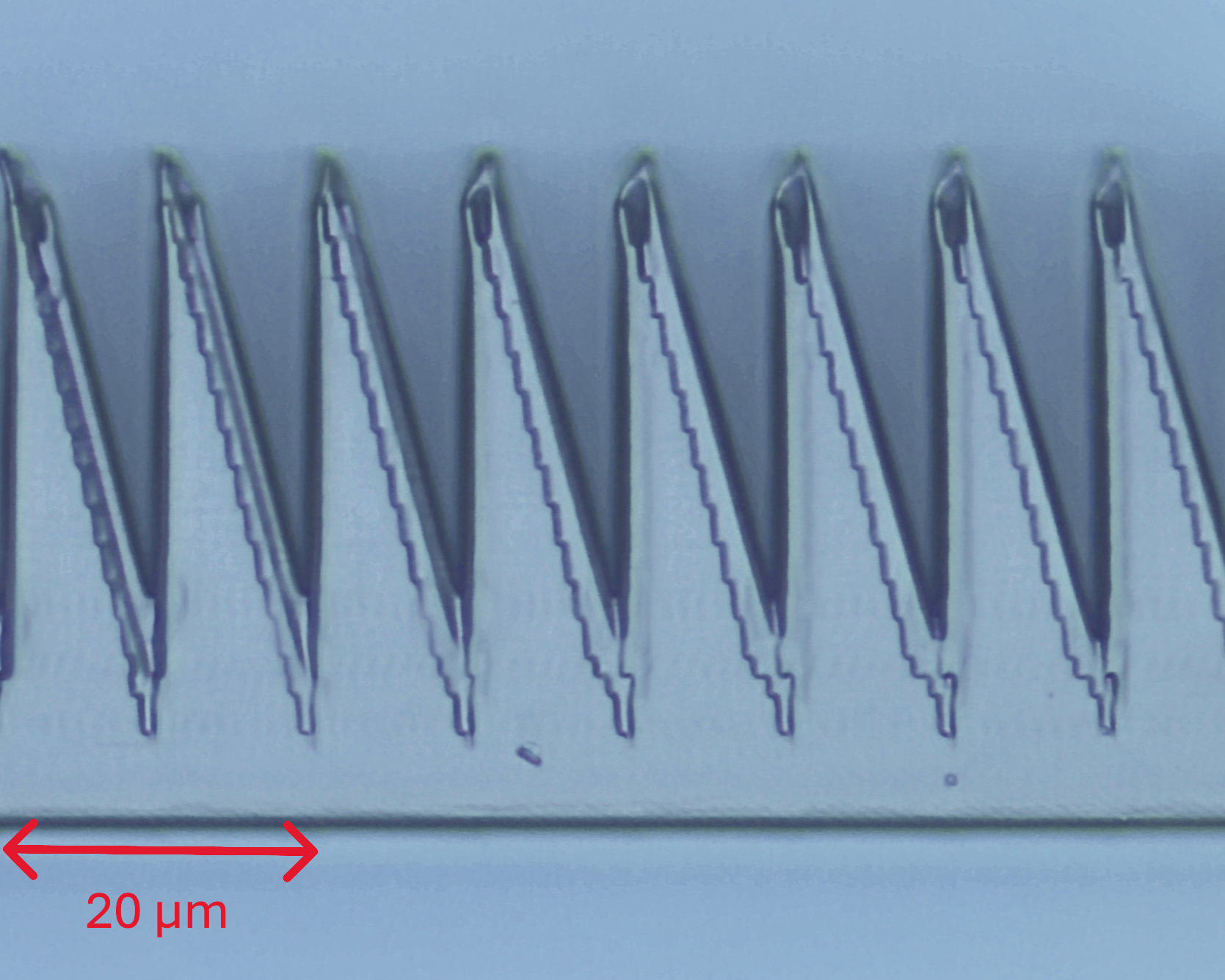}
    \caption{Optical image of parameter scan print, showing staircase structures and split prism tips due to overexposure.}
    \label{fig:pscan_steps_opt}
\end{figure}

 As previously mentioned, the size of the polymerization voxel needs to be tuned to the chosen slicing and hatching distances by adjusting the laser power and scan speed parameters. The goal is to have the shortest possible printing speed while delivering enough energy to polymerize the resin and concurrently avoid overexposure. Overexposure resulted in similar step-like structures as too large voxels, as can be seen in Figure~\ref{fig:pscan_steps_opt}, while underexposure can produce split or collapsed structures.

\begin{figure}
    \centering
    \includegraphics[width=0.9\linewidth]{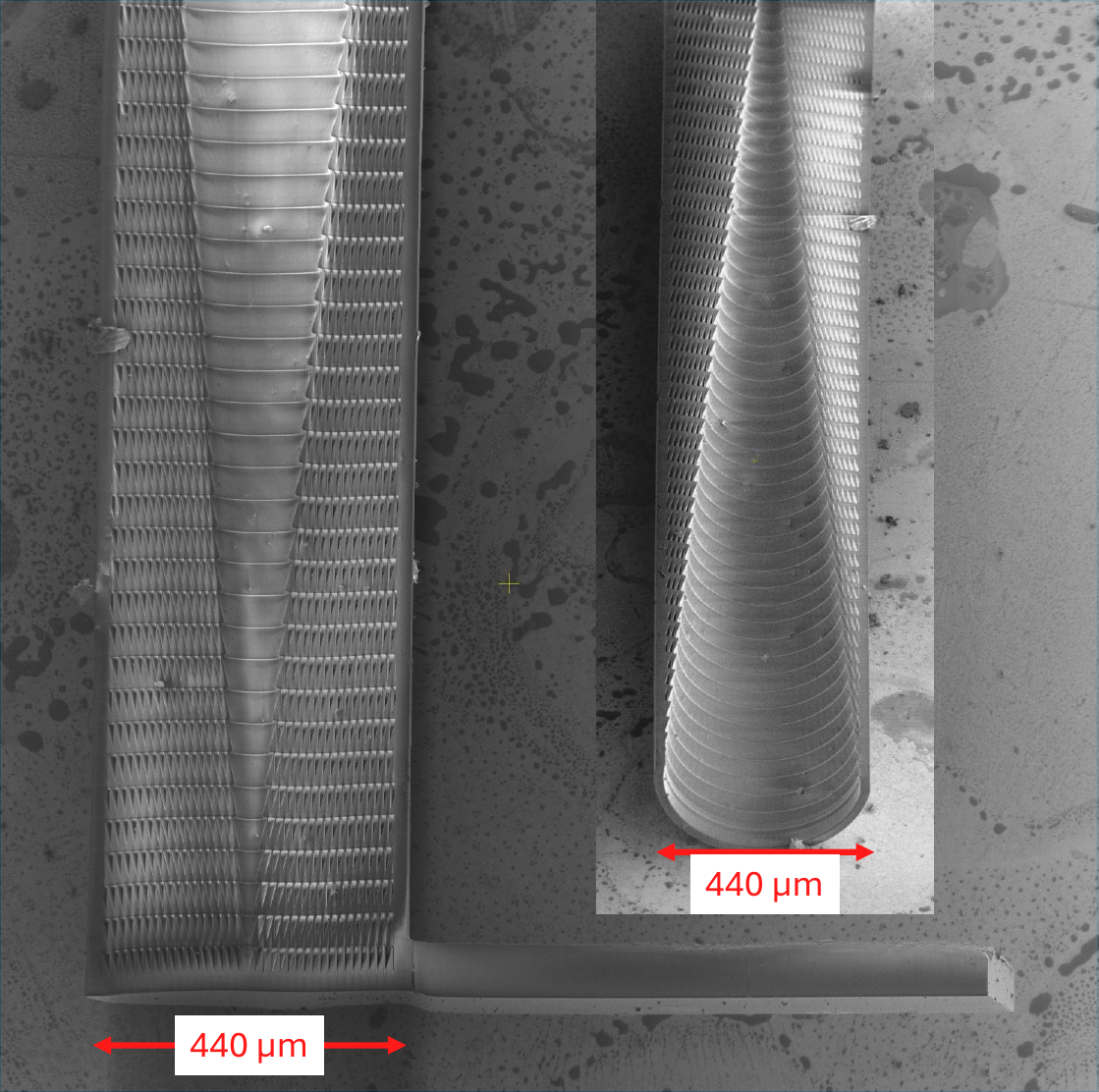}
    \caption{SEM images of 2PP manufactured lens, showing the bottom of the lens with the protruding tab (left), and the top of the lens at a 45 degree tilt (right).}
    \label{fig:full_lens_SEM}
\end{figure}

\begin{figure}
    \centering
    \includegraphics[width=0.5\linewidth]{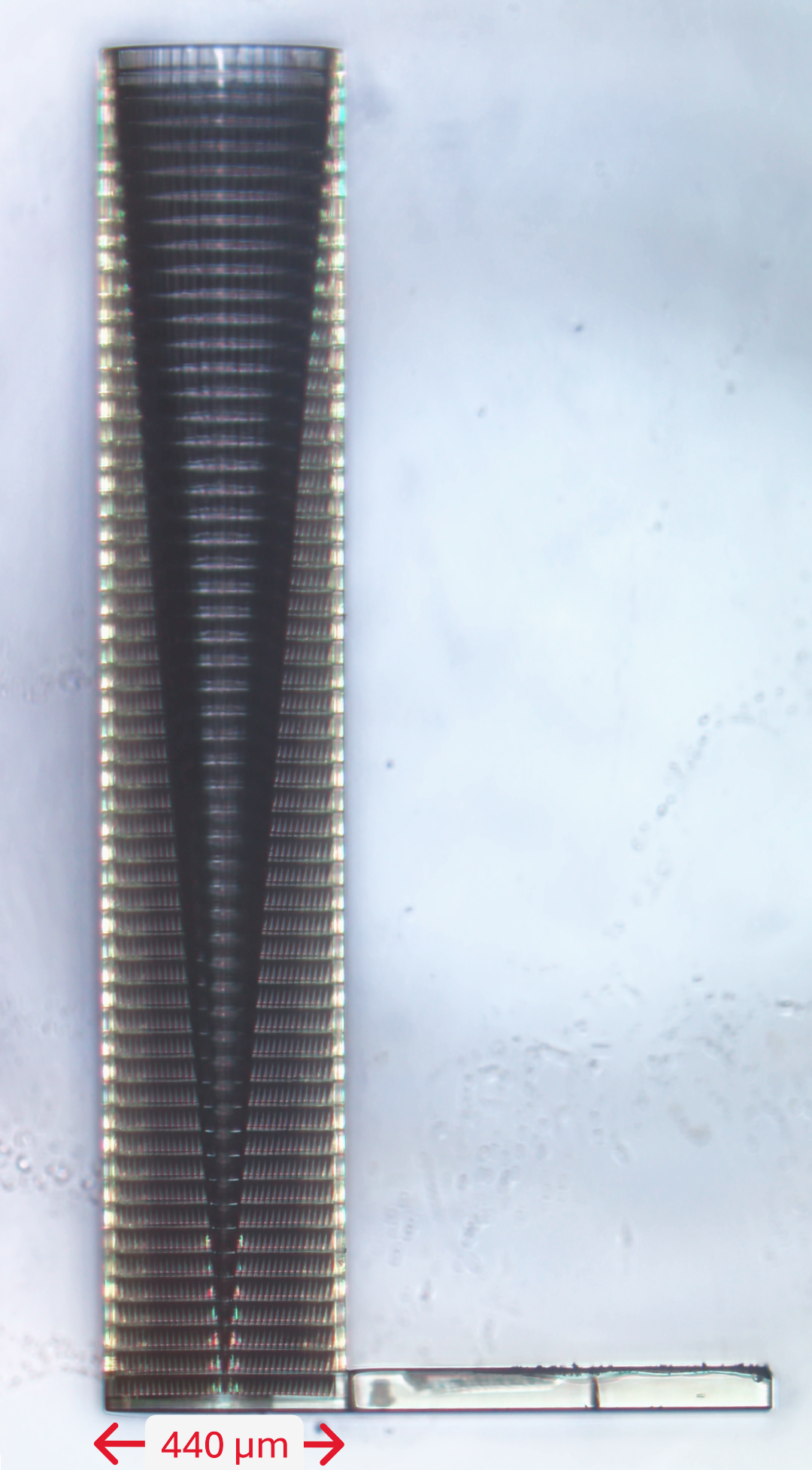}
    \caption{Optical microscope image of 2PP manufactured half-cylinder lens.}
    \label{fig:full_lens_optical}
\end{figure}

From the parameter sweep, laser power scaling set to 80\% and scan speed to 90 000~µm/s were found to give the smoothest surface and correct geometry with a reasonable compromise on printing speed. These parameters were then used to print a full lens in various configurations. SEM and optical images of a half-cylinder lens (for easier inspection) can be found in Figures~\ref{fig:full_lens_SEM} and~\ref{fig:full_lens_optical}. Ideally, a lens used to focus X-ray light would consist of a full cylinder, but a cutaway has to be included in the structure to provide an escape path for leftover resin after printing. If the lens is "closed" by printing the full 360 degree structure, resin will get trapped in the chambers between the prisms and  eventually cure, altering the prism structure and degrading the lens performance. Thus, a $30\degree$ cutaway was added in the CAD structure, which resulted in no liquid resin being present after the washing process. For a single SPL, the cutaway presents a defect in the point-spread function (PSF) and a loss of efficiency equal to the missing area. For a future SPL telescope which will utilize many SPL lenses taking parallel images, the first issue will be remedied by the cutaways having random orientations. The second issue will be reduced by making the cutaway as small as possible through a systematic study.

With the chosen lens design and printing parameters, the printing time for a full lens was 11~hours and 45~minutes. Since the aperture of such a lens is $\sim0.125$~mm$^2$, a future SPL telescope will need to utilize many SPL's to achieve a meaningful effective area (10's of thousands for a CubeSat mission, 100's of thousands for {\it XMM-Newton} or {\it Chandra}-sized missions). Thus,  the printing speed will need to be greatly improved upon in the future, and this will be discussed further in section~\ref{sec:conclusions}.

\section{Testing the 2PP SPL}\label{sec:testing_SPLs}

\subsection{Experimental setup}

\begin{figure}
    \centering
    \includegraphics[width=0.9\linewidth]{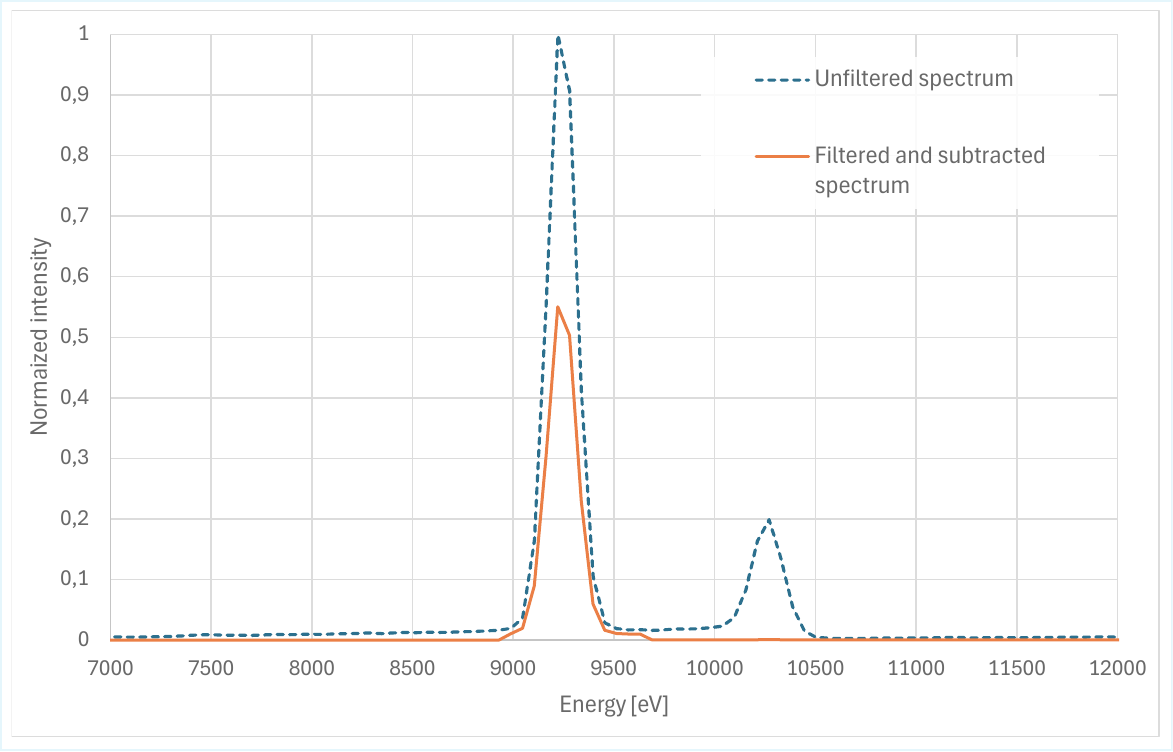}
    \caption{Part of the spectrum (showing the main fluorescence peaks) of the Excillum D2 source, before and after filtering and subtraction.}
    \label{fig:excillum_spectrum}
\end{figure}

To perform an initial characterization of the new SPL fabrication method, a test rig was setup in-house. An Excillum Metaljet D2 X-ray source was used, with an electron beam spot size of 20x70~µm and the acceleration voltage set to 50~kV. This produces an X-ray beam from a near-circular emission spot with a PSF width of $\sim20$~µm (FWHM). The central part of this spot is Gaussian but has significant tails in the transversal plane. The liquid metal alloy anode produces a diverging beam with a broad continuum emission and several fluorescence peaks, the most prominent of which are Ga K peaks at 9.2~keV and 10.3~keV. To isolate the 9.2~keV peak, two absorption-edge filters were used: a 20~µm Zn filter and a 17.5~µm Cu filter. By subtracting exposures for each separate filter, a narrow peak centered around the SPL design energy of 9.2~keV was produced, as can be seen in Figure~\ref{fig:excillum_spectrum}.

\begin{figure}
    \centering
    \includegraphics[width=0.98\linewidth]{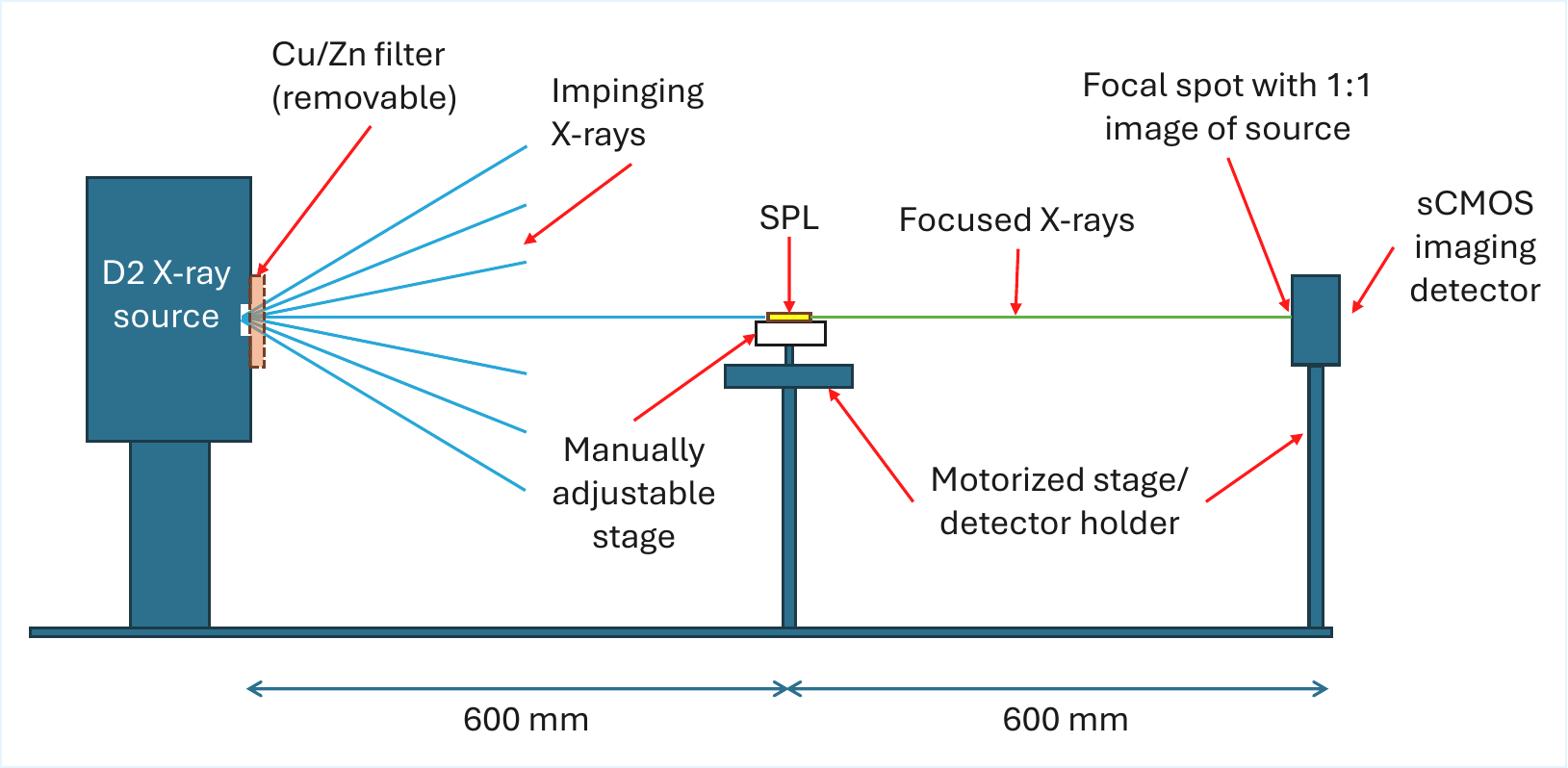}
    \caption{Schematic drawing of the test setup used to characterize the SPL.}
    \label{fig:test_schematic}
\end{figure}

The X-ray source was situated inside a shielded chamber with a maximum source-detector separation of 1348~mm. This allowed for a $4f$-setup with the 300~mm focal length SPL, with the detector and source both placed 600~mm from the lens. This creates a 1:1 scale image of the source at the detector plane. (Note that this is true for the focused light, but the projection of the lens geometry will be magnified by $2\times$.) The lens was placed in a 3D-printed holder affixed to a two-axis manually adjustable stage, which was then attached to the motorized stage of the X-ray chamber. This allowed for 3-axis movement and rotation along the axis perpendicular to the optical path. A Photonic Science sCMOS 16 MP imaging detector was used, with a pixel size of 9~µm and a reported PSF FWHM of 32~µm. A tungsten knife-edge was used to measure the combined FWHM of the source-detector system. The measured FWHM was 46 µm, in both the horizontal and vertical directions. If the X-ray source has the expected FWHM of $\sim20$~µm, this means that the detector has an actual FWHM of $\sim40$~µm. A schematic image of the test setup can be found in Figure~\ref{fig:test_schematic}.

After alignment, ten 20~second exposures were taken with each filter, and averaged before subtraction. The same exposures were then taken without the lens and stage present, to acquire a flat field for comparison. Similarly, a dark image was taken with the X-ray source off for background subtraction.

\subsection{Results}

\begin{figure}
    \centering
    \includegraphics[width=0.9\linewidth]{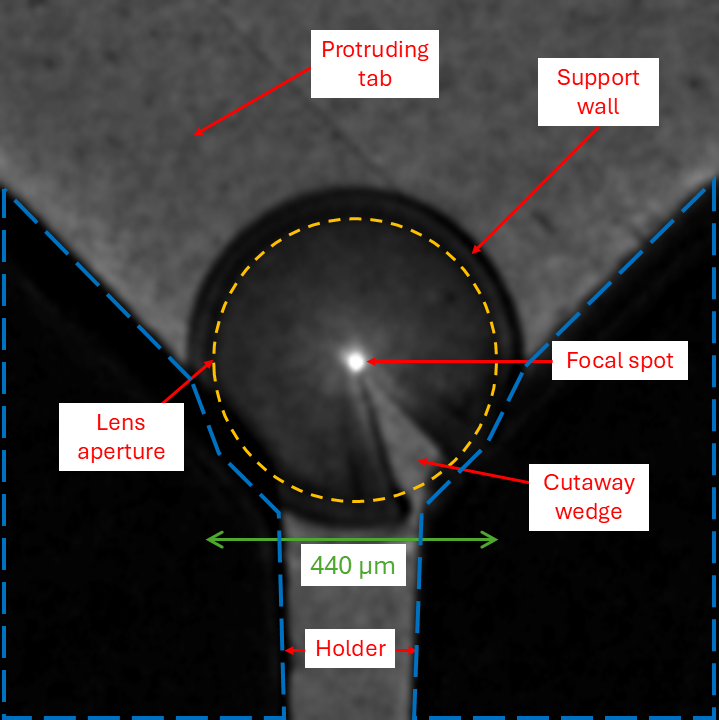}
    \caption{X-ray image of an SPL focusing light. The yellow circle corresponds to the working lens aperture with a diameter of 400~µm, while the full lens including the support walls has a diameter of 440~µm. The protruding tab used for handling the lens can be seen as a rectangular shadow in the top left. The part of the plastic holder visible in the image is outlined in blue.}
    \label{fig:focusing_light}
\end{figure}

The acquired images were processed and inspected using Fiji\cite{schindelin_fiji_2012}. After averaging each filtered exposure and subtracting them, the image in Figure~\ref{fig:focusing_light} was obtained. The X-ray light is focused to a central point, with the rest of a the lens aperture being significantly darker than the flat field beside the lens.

\begin{figure}
    \centering
    \includegraphics[width=0.9\linewidth]{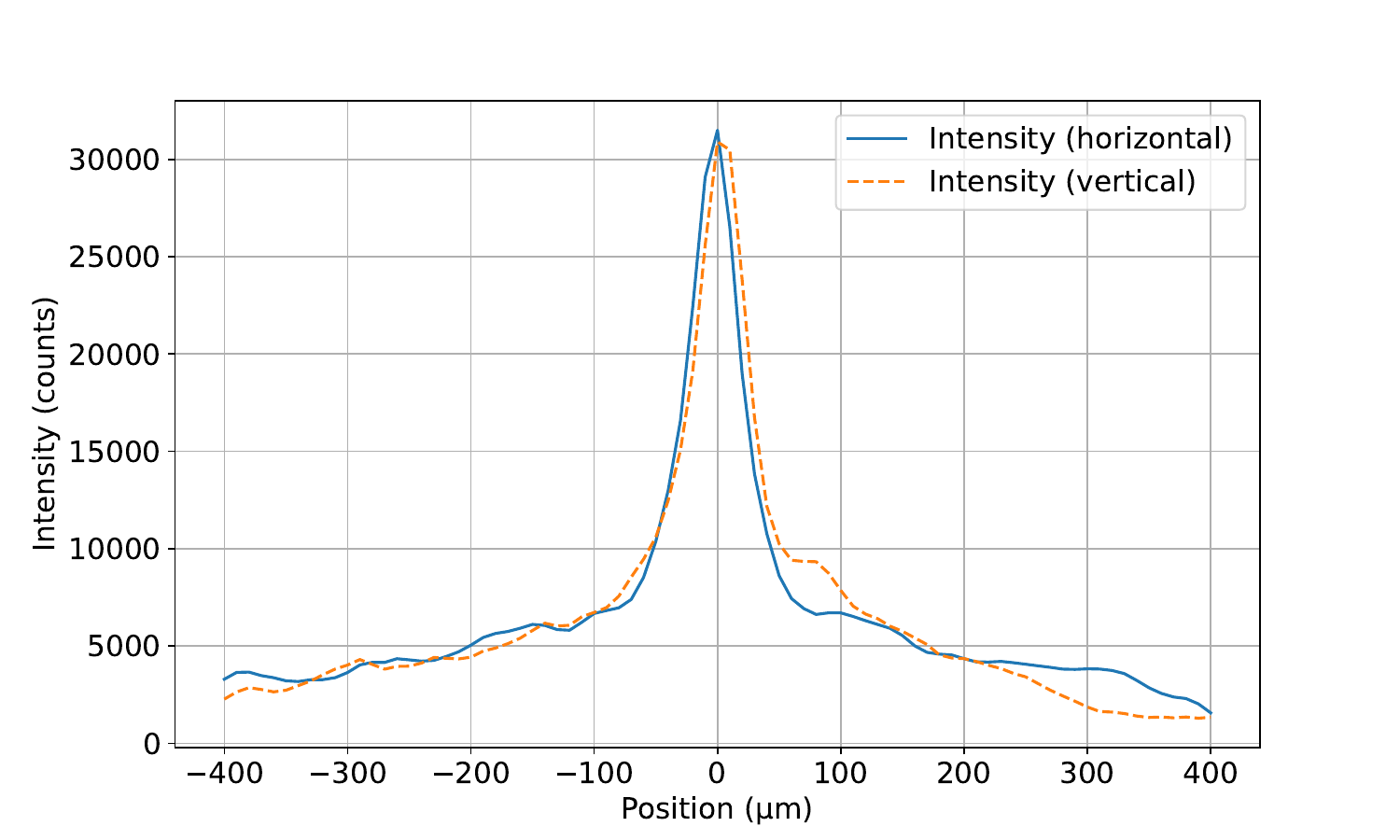}
    \caption{Intensity profile across the aperture in the horizontal and vertical directions. The central peak has a Gaussian FWHM of 49.3~µm (horizontal) and 47.2~µm (vertical).}
    \label{fig:v/h_profile}
\end{figure}

To characterize the focal properties of the SPL, the images were analyzed with a Python script, where horizontal and vertical slices across the aperture were extracted, producing the profiles of Figure~\ref{fig:v/h_profile}. The central peak is Gaussian with a FWHM of 49.3~µm (horizontal) and 47.2~µm (vertical). Aside from this peak, the expected broad tails from the X-ray beam can be seen. The FWHM of the Gaussian peak is similar to the measured combined FWHM of the detector and X-ray beam, which indicates that the FWHM of the SPL is lower than the error margin on this measurement. The error margin is dominated by the uncertainty of the measured FWHM from the source-detector system. If this uncertainty is taken to be half the physical width of one pixel (4.5~µm), then the largest possible FWHM of the lens PSF is 34.2~µm. This corresponds to the case where the actual FWHM of the source-detector system is 41.5~µm, and the real FWHM with the lens is 53.8~µm. This would give a FWHM contribution of the SPL as $\sqrt{53.8^2-41.5^2} = 34.2$~µm, assuming $\sigma_\mathrm{measured}^2 = \sigma_\mathrm{system}^2 + \sigma_\mathrm{lens}^2$. Converting 34.2~µm to angular resolution gives 23.5 arcseconds, which represents the upper bound on the SPL angular resolution. This is a conservative estimate, and the actual resolution of the SPL will need to measured using more specialized equipment.

To measure the efficiency of the lens, a circular region equal to the aperture was extracted from the image, as shown in Figure~\ref{fig:focusing_light}. This aperture is affected by the support wall, due to edge effects partially stemming from the diverging beam, imperfect alignment and detector resolution. To avoid this, the aperture was defined to be slightly smaller than its radius of 200~µm, i.e. 44 pixels (since at 2x magnification, the apparent pixel size is 4.5~µm, and $200/4.5 \approx 44$), and efficiency calculations were also performed for this "functional aperture" with a radius of 40 pixels. This corresponds to a smaller SPL with a radius of 182.25~µm.

The total number of counts inside the aperture was integrated for the lens image and the flat field exposure for both defined apertures. To exclude the area where lens material is not present due to the $30\degree$ cutaway in the lens, 1/12 of the intensity of the flat field exposure was subtracted from both integrated intensities. Then, the lens intensity was divided with the flat field intensity to get the efficiency, and the effective aperture (defined as $A_\mathrm{eff} = A\sqrt{\epsilon}$, where $A$ is the geometric aperture and $\epsilon$ is the efficiency). The results compared with the analytical efficiency (as calculated from Equation~\ref{eq:transmission_func}) can be found in Table~\ref{tab:efficiency}, which also contains a comparison to the UV lithography (UVL) SPL manufactured and tested by Ref.~\citenum{mi_stacked_2019}.

The apparent lower efficiency of the 2PP SPL compared to Ref.~\citenum{mi_stacked_2019} is due to the shorter focal length (0.3~m instead of 0.7~m) and lower operational energy (9.2~keV instead of 13.5~keV) both of which decreases the efficiency as described in Section~\ref{sec:optimizing_spl}. However, the 2PP SPL efficiency is closer to analytical values than the UVL SPL, indicating a better performance in terms of efficiency for the actual manufactured lens. As described in Section \ref{sec:SPL_design} an SPL can be designed for any X-ray energy and focal length, and it can reasonably be assumed that a 2PP SPL designed for 13.5~keV and a focal length of 0.7~m would thus have an improved efficiency over the UVL SPL, given the results in Table \ref{tab:efficiency}.

\begin{table}[h]
    \centering
    \setlength{\extrarowsep}{4pt}
    \begin{tabu}{X[1.2,c] X[c] X[1.3,c] X[1.3,c] X[c] X[2.2, c] X[1, c]}
        \hline\hline
         Lens & Geometric aperture & Efficiency (analytical) & Efficiency (measured)  & Effective aperture  & Effective aperture/ geometric aperture & Effective area\\
            & [µm] & & & [µm] & & [mm$^2$]\\
            \hline 
         2PP SPL & $364.5$ & $66.3\%$ & $61.0^{+1.8}_{-0.7}\%$ & $285^{+4}_{-2}$ & $78.2^{+1.1}_{-0.6}\%$ & $0.0638^{+0.018}_{-0.009}$\\
         2PP SPL & $400$ & $63.6\%$ & $53.7^{+1.9}_{-1.6}\%$& $322^{+6}_{-4}$ & $73.3^{+1.3}_{-1.1}\%$ & $0.0814^{+0.018}_{-0.009}$ \\
         UVL SPL & $420$ & $86.7\%$ & $70\%$ & $352$ & $84\%$ & $0.0973$\\ 

        \hline
    \end{tabu}
    \caption{Efficiency, effective aperture and effective area of single* 2PP SPL using the functional aperture and the full geometric aperture, compared with the UV lithography (UVL) SPL previously tested\cite{mi_stacked_2019}.\\
    *An SPL telescope, as discussed in this article, would consists of $10^4-10^6$ SPLs.}
    \label{tab:efficiency}
\end{table}

\section{Conclusions and Outlook}\label{sec:conclusions}

This paper shows that two-photon polymerization manufacturing of SPLs improves on previous manufacturing methods in terms of geometric fidelity and production time. The manufactured lens has better efficiency and a FWHM resolution smaller than 23.5~arcseconds. The actual resolution is expected to be at the sub-arcsecond level, with precise measurements to be performed at a later date. These factors indicates that 2PP manufacturing is a viable path forward for the SPL concept, and provides an important step towards increasing the technology-readiness level for a future SPL satellite telescope. 2PP printing with commercially-available machines such as the Nanoscribe Photonic Professional GT2 can produce a working lens with high levels of geometric fidelity. The ability to print any geometry and produce a 55-layer working lens on a timescale of 10 hours provides a vast improvement on the methods used by Ref.~\citenum{mi_stacked_2019}, where a 30-layer lens took more than 200 hours to produce. Nevertheless, the printing time will still need to be further optimized to make an SPL array telescope feasible. 

To fill a $10\times10$~cm aperture (representative for a small proof-of-concept CubeSat SPL telescope), 10's of thousands of 200~µm-radius SPLs would be needed. This means that the current printing time will need to be reduced by several orders of magnitude. Several possible avenues exist to achieve this reduction. The first is to (somewhat counter-intuitively) reduce the radius of each individual SPL. Since a larger radius necessitates more layers and therefore a longer lens, the total polymer volume to be manufactured scales as $R^3$. The printing time is roughly proportional to this volume, and since the geometric area of the aperture scales as $R^2$ a linear gain in printing time can be attained by reducing the radius and increasing the number of lenses in the array. As is shown in Figure~\ref{fig:tradeoff_9.2kev}, reducing the radius also improves the efficiency of the telescope. Reducing the radius does come with its own challenges in terms of an SPL telescope, however. Assembly, collimation and detector systems increase in complexity with the number of lenses, even if the SPLs are printed many at a time in sub-arrays. Pick-and-place solutions from the microelectronics industry are likely to provide solutions here.

Secondly, the printing time can be reduced by using a larger objective for the 2PP machine. Nanoscribe offers a 10x objective for their machines, which allows for faster printing by a factor of $\sim 50-100$ compared to the 25x. The larger printing field (diameter 1~mm) would also allow for four $R=200$~µm SPLs to be printed at once. The drawback naturally comes in terms of print fidelity; a larger objective cannot print as fine structures, and typical slicing and hatching distances for the 10x are 2-5 times those of the 25x. This would mean a significant drop in SPL print quality, and a systematic study on the effect on the lens performance would be needed.

Lastly, the quickly developing field of 2PP 3D printing and nanomanufacturing in general could provide opportunities for improvement in terms of printing speed. The machine used for this work was acquired in 2018, and has been superseded as Nanoscribe's flagship model by the {\it Quantum X} line. These new machines utilize two-photon grayscale lithography for 3D printing, quoted by the company as improving printing time by a factor of 10 to 60 compared to traditional 2PP\cite{nanoscribe_quantumX}, while simultaneously improving surface smoothness. Such a machine or similar systems produced by other companies is a promising avenue for improvement. The method is also inherently scalable in terms of parallelizing production across several machines, and further performance improvements can be expected going forward as the 2PP process is a well-established research and industrial tool.

Even with the most conservative estimates, these factors improve SPL printing time by three orders of magnitude, meaning that over 1000 lenses could be manufactured in the 12 hours needed for the single lens in this paper. This means that printing 10's of thousands of lenses for an SPL telescope is within the realm of feasibility. An important future study will be to optimize the printing process to achieve the best printing fidelity for SPL mass production.

To further study the performance of 2PP-manufactured SPLs and quantify the angular resolution, tests at a synchrotron light source (similar to those performed by Ref.~\citenum{mi_stacked_2019}) will need to be performed. Several beamlines exist with spot sizes and detector resolution small enough to measure a possible sub-micron resolution in terms of FWHM, and measure the HPD (half-power diameter) and intensity gain in the focal spot. The off-axis vignetting will also be studied, and the SPL response to a non-monoenergetic beam, characterizing the chromacity. Additionally, comparisons in terms of performance for different printing parameters will be made, to further assist in optimizing and speeding up the printing process. Synchrotron tests are currently planned for Spring/Summer 2026.

Thermal vacuum (TVAC) and structural testing is required to confirm that the polymer structure of 2PP printed SPL can survive a launch and subsequent operation in a space environment. The SEM study performed for this work was performed in a $10^{-5}$~mbar vacuum and no indication of structural damage was seen on the SPLs, indicating short-term tolerance of a vacuum environment.

An SPL telescope will likely be based on the concept shown by Ref.~\citenum{mi_stacked_2019}. Realizing such a telescope will mean overcoming several challenges, which are outlined here and will be the subject of future work. As described in Ref.~\citenum{mi_stacked_2019}, the 10's of thousands of SPLs will each have their own collimators and share a detector plane. The spatial resolution these detectors need to fully take advantage of the strengths of SPLs have yet to be realized, but promising steps have been taken in the last few years\cite{brunskog_experimental_2024}. Detector planes could also be stacked to capture light focused at different distances, or different parts of the telescope aperture could have SPLs tuned to different energies. Even with these corrections, chromatic aberration is the main drawback of refractive optics, and the choice of SPL energies would need to be made carefully to achieve an acceptable energy response and ensure scientific relevance. It should be noted that unfocused energies could still be measured by the detector, and depending on the objects in the field of view this could provide some broadband spectral information. Here the high efficiency of SPLs provide a major strength, opening the possibility of a combined imaging/non-imaging telescope. The images from all SPLs in a telescope will be stacked to form a single image. This will be a non-trivial process with strong requirements on background reduction.

For the assembly of 10's of thousands of SPLs into a telescope system, methods from the microelectronics industry are likely the best choice. Regardless of the final manufacturing method used, SPLs could be printed together in small arrays (necessitating a clever solution for the collimation) which could be then combined into larger arrays, minimizing handling. These challenges will need to be addressed in conjunction with the choice of manufacturing method, to be discussed in a future work.

\section*{Disclosures}

The authors declare there are no financial interests, commercial affiliations, or other potential conflicts of interest that have influenced the objectivity of this research or the writing of this paper.

\section*{Code and Data Availability}
The code and data used to produce the results of this paper can be obtained by request from the authors.

\section*{Acknowledgments}

The funding for this work was provided by the Swedish National Space Agency (grant number 2023-00308), through the Technology Research for Space Applications Programme. The authors would like to thank Niclas Roxhed (KTH) for facilitating the 2PP printing process, the staff at the MST Lab and AlbaNova Nanolab at KTH for support during manufacturing and SEM studies, respectively, Jakob Larsson and Nirmal Iyer at Excillum AB for assistance in testing the SPL, and Mats Danielsson (KTH), Wujun Mi and Peter Nillius for guidance and background on the SPL concept.

No generative AI tools were used to produce any of the text or figures of this paper. An AI tool (GitHub Copilot) was used to assist in the programming of several Python scripts, used for analytical calculations and data analysis.

%%%%% References %%%%%

\bibliography{report.bib}   % bibliography data in report.bib
\bibliographystyle{spiejour.bst}   % makes bibtex use spiejour.bst

%%%%% Biographies of authors %%%%%

\vspace{2ex}\noindent\textbf{Filip af Malmborg} is a PhD student at KTH Royal Institute of Technology and the Oskar Klein Centre for Cosmoparticle Physics in Stockholm, Sweden. He received his Bachelor's and Master's degrees in Engineering Physics with specialization in Subatomic Physics and Astrophysics in 2020 and 2023, respectively. He has previously worked on the XL-Calibur X-ray polarimetry telescope. His research interests include X-ray telescope design, X-ray astronomy, high-energy astrophysics, extreme objects and computer simulation.

\vspace{1ex}
\noindent Biographies of the other authors are not available.

\listoffigures
\listoftables

\end{spacing}
\end{document}